\documentclass[twocolumn,showpacs,preprintnumbers,amsmath,amssymb]{revtex4}

\usepackage{graphicx}
\usepackage{dcolumn}
\usepackage{bm}

\begin{document} 

\preprint{}

\title{Kondo effect in a semiconductor quantum dot \\coupled to ferromagnetic electrodes}

\author{K. Hamaya,\footnote{hamaya@iis.u-tokyo.ac.jp}\footnote{Also at: Institute for Nano Quantum Information Electronics, University of Tokyo.} M. Kitabatake, K. Shibata, M. Jung, M. Kawamura, K. Hirakawa,$^{\dag}$\footnote{Also at: Japan Science and Technology Agency, CREST.} and T. Machida$^{\dag}$$^{\ddag}$\footnote{tmachida@iis.u-tokyo.ac.jp}}
\affiliation{%
Institute of Industrial Science, University of Tokyo, 4-6-1 Komaba, Meguro-ku, Tokyo 153-8505, Japan
}%

\author{T. Taniyama\footnote{Also at: Japan Science and Technology Agency, PRESTO.}}
\affiliation{%
Materials and Structures Laboratory, Tokyo Institute of Technology, 4259 Nagatsuta, Midori-ku, Yokohama 226-8503, Japan 
}%

\author{S. Ishida and Y. Arakawa$^{\dag}$}
\affiliation{%
Research Center for Advanced Science and Technology and Institute of Industrial Science, University of Tokyo, 4-6-1 Komaba, Meguro-ku, Tokyo 153-8505, Japan }

%

\date{\today}
\begin{abstract}
Using a laterally-fabricated quantum-dot (QD) spin-valve device, we experimentally study the Kondo effect in the electron transport through a semiconductor QD with an odd number of electrons ($N$). In a parallel magnetic configuration of the ferromagnetic electrodes, the Kondo resonance at $N =$ 3 splits clearly without external magnetic fields. With applying magnetic fields ($B$), the splitting is gradually reduced, and then the Kondo effect is almost restored at $B =$ 1.2 T. This means that, in the Kondo regime, an inverse effective magnetic field of $B \sim$ 1.2 T can be applied to the QD in the parallel magnetic configuration of the ferromagnetic electrodes.
\end{abstract}
\maketitle
Using fabrication techniques of nano-gap electrodes consisting of Ti/Au, Jung {\it et al.} established lateral transport measurements of electrons through a single self-assembled InAs quantum dot (QD),\cite{Jung1} and observed shell-dependent charging energies due to an artificial atomic nature.\cite{Jung2} Recently, Shibata {\it et al.} and Igarashi {\it et al.} got a device with strong coupling between the InAs QD and Ti/Au electrodes and succeeded in the observation of the Kondo effect in a single self-assembled InAs QD.\cite{Shibata,Igarashi} The Kondo effect, which is due to the screening of a localized spin in the QD from conduction electrons in the electrodes, is one of the many-body phenomena.\cite{Gold,Cronenwett,Sasaki} Below the Kondo temperature, the zero-bias conductance is enhanced through spin-flip processes, i.e., Kondo resonance. In external magnetic fields, the Kondo effect in QDs with an odd number of electrons is suppressed due to the spin splitting of the unparied localized electron state, resulting in splitting of the Kondo resonance.\cite{Gold,Cronenwett,Sasaki} 

On the other hand, using similar fabrication techniques, we have fabricated lateral QD spin-valve (SV) devices which consist of a single self-assembled InAs QD with ferromagnetic electrodes. To date, clear spin transport via the tunneling magnetoresistance (TMR) effect was observed in Co/InAs QD/Co and Ni/InAs QD/Ni QDSVs.\cite{Hamaya1,Hamaya2} Since we can achieve an electric-field control of the TMR,\cite{Hamaya1,Hamaya2} these QDSVs are expected to be a novel spintronic application such as a gate-tunable spin memory.\cite{Datta,Sahoo} Recently, the Kondo effect in the presence of ferromagnetism was studied in Ni/C$_{60}$/Ni SVs,\cite{Ralph} and the splitting of the Kondo resonance (suppression of the Kondo effect) was demonstrated even in the absence of external magnetic fields. However, the Kondo effect and its related phenomena for semiconductor QDSVs have never been reported.
\begin{figure}[t]
\includegraphics[width=8cm]{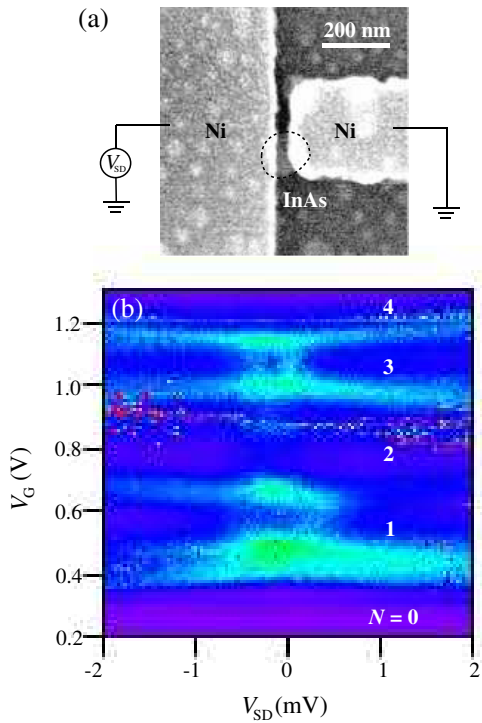}
\caption{(Color online) (a) A scanning electron micrograph of our Ni/InAs/Ni QDSV device. (b) A color scale plot of differential conductance, $dI$/$d$$V$$_\mathrm{SD}$, as a function of $V$$_\mathrm{SD}$ and $V$$_\mathrm{G}$ without an external magnetic field. Light regions show the finite $dI$/$d$$V$$_\mathrm{SD}$ regime and the number of electrons is indicated.}
\end{figure}  

In this Letter, we experimentally study the Kondo effect in the electron transport through a semiconductor QD with an odd number of electrons ($N$) using a Ni/InAs QD/Ni QDSV device. Not only suppression of the Kondo effect but also its restoration by applying external magnetic fields ($B$) is demonstrated. In a parallel magnetic configuration of the Ni electrodes, the zero-bias anomaly at $N =$ 3 splits clearly without an external magnetic field. With applying magnetic fields, the splitting gradually disappears, and then the Kondo effect is almost restored at $B =$ 1.2 T. This fact indicates that an inverse effective magnetic field of $B \sim$ 1.2 T can be applied to the QD in the parallel magnetic configuration of the ferromagnetic electrodes.
 
We grew self-assembled InAs QDs on a substrate consisting of 170-nm-thick GaAs buffer layer/90-nm-thick AlGaAs insulating layer/$n$$^{+}$-GaAs(001). The $n$$^{+}$-GaAs(001) was used as a backgate electrode. The wire-shaped Ni electrodes with a $\sim$ 30-nm gap were fabricated by using a conventional electron-beam lithography and a lift-off method. Before the evaporation of Ni thin films, we etched InAs surface with buffered HF solution for 6 s. The detailed fabrication procedures have been described in our previous works.\cite{Hamaya1,Hamaya2} An enlarged scanning electron micrograph of the QDSV used here is shown in Fig. 1(a). We can see a single InAs QD (120 $\sim$ 140 nm) in contact with two Ni electrodes which have a thickness of 40 nm. The overlapped regions of the InAs QD reach more than $\sim$ 30 nm. Before cooling down of the device, we have checked the two-terminal contact resistance of $\sim$ 70 k$\Omega$ at room temperature. We emphasize that Ti seed layer is not used, which is unlike the case of previous works.\cite{Shibata,Igarashi} Transport measurements were performed by a dc method at 30 mK. External magnetic fields ($B$) were applied parallel to the long axis of the wire-shaped Ni electrodes. 
\begin{figure}[t]
\includegraphics[width=8.5cm]{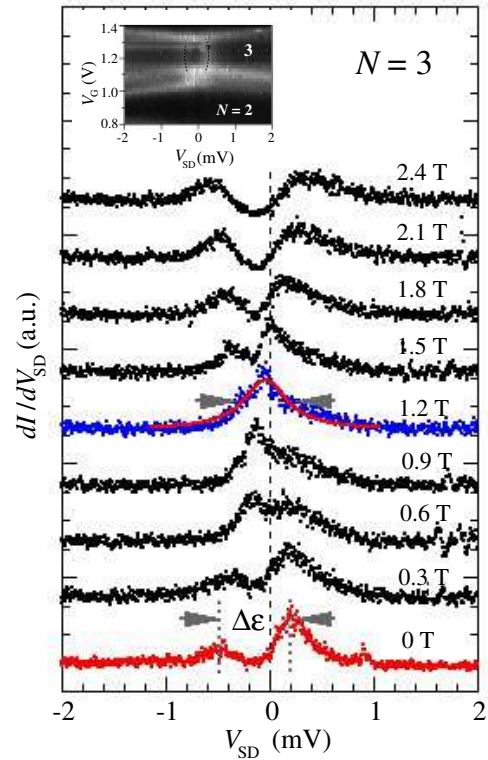}
\caption{(Color online) The $dI$/$d$$V$$_\mathrm{SD}$ as a function of $V$$_\mathrm{SD}$ for various external magnetic fields at $V$$_\mathrm{G} =$ 1.25 V ($N =$ 3). The inset shows the corresponding Coulomb diamonds near $N =$ 2 and 3.}
\end{figure} 

First of all, we applied a magnetic field of $B =$ 1 T to the wire-shaped Ni electrodes and made a parallel magnetic domain configuration. After the saturation of the magnetization, when the magnetic field is reduced down to $B =$ 0 T, the magnetic configuration remains parallel because of the strong shape anisotropy of the Ni electrodes.\cite{Hamaya1,Hamaya2} In this condition, we measure the differential conductance $dI$/$d$$V$$_\mathrm{SD}$ as a function of $V$$_\mathrm{SD}$ and $V$$_\mathrm{G}$. Figure 1(b) displays a representative data; The light regions exhibit higher conductance regimes and the number of electrons ($N$) is indicated. We can see the enhancement in the conductance in the vicinity of zero bias ($V$$_\mathrm{SD} \sim$ 0 V) at $N = $1 and 3 Coulomb valleys. The enhanced conductance at $N = $1 and 3 implies the $S =$ 1/2 Kondo effect, as previously reported by Shibata {\it et al.} and Igarashi {\it et al}.\cite{Shibata,Igarashi} Surprisingly, we observe an evident splitting of the zero-bias anomaly at $N =$ 3,\cite{ref} i.e., suppression of the Kondo effect at $V$$_\mathrm{SD} =$ 0 V, without an external magnetic field ($B =$ 0 T). The $dI$/$d$$V$$_\mathrm{SD}$ vs $V$$_\mathrm{SD}$ measured at $N =$ 3 is shown in the red curve in Fig. 2.\cite{ref2} The peak spacing ($\Delta$$\epsilon$) of the splitting is $\sim$ 0.74 mV. The splitting feature shows an asymmetry about the polarity of $V$$_\mathrm{SD}$. This feature may result from the asymmetric tunnel coupling between ferromagnetic electrodes and the QD,\cite{Utsumi} which is not controllable for these systems,\cite{Jung1,Jung2,Shibata,Igarashi,Hamaya1,Hamaya2} or the difference in the spin polarization at the Ni/InAs interfaces.\cite{Utsumi} 

With increasing external magnetic fields parallel to the long axes of the wire-shaped electrodes, the splitting of the Kondo resonance is gradually reduced, and the observed double peak changes into a single peak at $B =$ 1.2 T (blue curve in Fig. 2). This means that the Kondo effect is almost restored. From the $\Delta$$V$$_\mathrm{SD}$ of the Kondo resonance measured at $B =$ 1.2 T, we can roughly estimate the Kondo temperature $T_\mathrm{K}$ of $\sim$ 5 K, which is consistent with that of Au/Ti/InAs QD/Ti/Au systems.\cite{Shibata,Igarashi} For further increase in applied magnetic fields, the Kondo effect is suppressed again due to the lifting of the spin degeneracy. 

To examine the relationship between the zero-bias conductance and the magnetic configuration of Ni electrodes, we also investigate the magnetoconductance of the QDSV at $V$$_\mathrm{G} =$ 1.25 V ($N =$ 3) in Fig. 3. The red and blue curves correspond to the data for up-sweep and down-sweep measurements, respectively. At $\pm$ $\sim$ 0.015 T we can see current changes due to the switching of the magnetic configuration of the Ni electrodes from parallel to anti-parallel:\cite{Hamaya2} we can see an enhancement in the conductance for anti-parallel configurations, i.e., inverse TMR effect. This means that the Kondo effect is suppressed in a parallel magnetic configuration in the Ni electrodes, while the Kondo effect is somewhat restored in an anti-parallel magnetic configuration.\cite{Martinek1,Martinek2,Utsumi,Ralph} From these data, we regard the suppression of the $N =$ 3 Kondo effect at $B =$ 0 T as a consequence of the ferromagnetic electrodes with a parallel magnetic configuration. 
\begin{figure}[t]
\includegraphics[width=8.5cm]{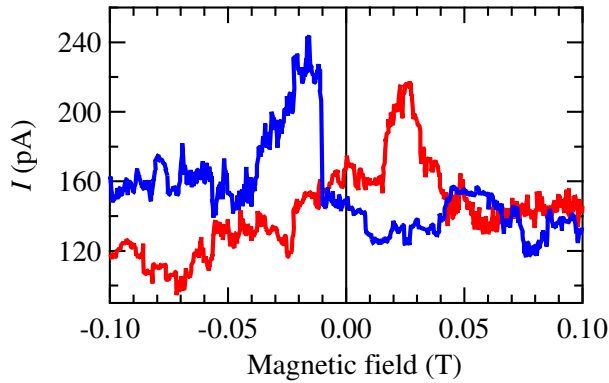}
\caption{(Color online) The magnetoconductance curve measured at $V$$_\mathrm{G} =$ 1.25 V ($N =$ 3).}
\end{figure} 

In Au(Co)/Au QD/Au(Co) metallic junctions, suppression of the Kondo effect at $B =$ 0 T has also been observed.\cite{Heersche} The splitting of the Kondo resonance results from the competition between the Kondo effect and the indirect RKKY interaction.\cite{Heersche} However, it is known that the RKKY-like interaction between the spin of the QD and the static spins of the Co impurities can not induce evident hysteretic TMR curves,\cite{Heersche} as shown in Fig. 3. Accordingly, we can rule out the RKKY interaction between the QD spin and the static spins of the Ni electrodes. Next, we in turn need to consider an influence of a local stray field from the ferromagnetic electrodes. When the magnetic configuration of the ferromagnetic electrodes is parallel, the local stray field can induce the  Zeeman splitting, where the direction of the local stray field can be regarded as the same direction of external magnetic fields. However, Fig. 2 clearly indicates that an effective field inducing the suppression of the Kondo effect works in the opposite direction of external magnetic fields. Thus, the splitting of the Kondo resonance presented here can not be explained by the local stray field. 

From the above considerations, we now infer that the suppression of the Kondo effect at $B =$ 0 T is probably due to a local exchange field, which is similar to Ni/C$_{60}$/Ni SVs previously shown.\cite{Ralph} In theories,\cite{Martinek1,Martinek2,Utsumi} for parallel magnetic configurations of ferromagnetic electrodes in QDSV devices, the spin-dependent quantum charge fluctuations occur, and then the QD's levels are renormalized. The renormalization for the spin-down level is stronger than that for the spin-up level, so that a level splitting between the two spin orientations is generated.\cite{Martinek1,Martinek2,Utsumi} It has been well discussed that such effective field is compensated by applying magnetic fields,\cite{Martinek1,Martinek2,Utsumi} being consistent with the present data in Fig. 2. For previous Ni/C$_{60}$/Ni SVs,\cite{Ralph} since the exchange field probably acted as an effective field parallel to the magnetization orientation of the Ni electrodes, restoration of the Kondo effect by applying external magnetic fields could not be detected. In the present case, the exchange field works as an effective field anti-parallel to the the magnetization orientation of the Ni electrodes, so that the restoration of the Kondo effect can be observed clearly by applying external magnetic fields, as illustrated in Fig. 2. 

In summary, we have studied the Kondo effect using a laterally-fabricated semiconductor QDSV device. In a parallel magnetic configuration of the ferromagnetic electrodes, the Kondo resonance at $N =$ 3 splits clearly in the absence of external magnetic fields. With applying magnetic fields, the splitting is gradually reduced, and then the Kondo effect is almost restored at $B =$ 1.2 T. This means that, in the Kondo regime, an inverse effective magnetic field of $B \sim$ 1.2 T can be applied to the QD in the parallel magnetic configuration of the ferromagnetic electrodes.
 
K. H. acknowledges Y. Utsumi for his fruitful discussions. This work is supported by the Special Coordination Funds for Promoting Science and Technology, and the Grant-in-Aid from MEXT, and Collaborative Research Project of Materials and Structures Laboratory, Tokyo Institute of Technology, and the Sumitomo Foundation.


\begin{thebibliography}{11}
\bibitem{ref}
The zero-bias anomaly at $N =$ 1 is also likely to be split but the splitting is not so clear. Thus, we focus on the data at $N =$ 3.
\bibitem{ref2}
Unfortunately, a small change in gate properties was brought about, but the splitting of the zero-bias anomaly at $N =$ 3 was reproduced, as shown in the inset of Fig. 2. Hence the data presented in Fig. 2 were measured at $V$$_\mathrm{G} =$ 1.25 V in the inset figure.
\end{thebibliography}
\end{document}